\documentclass[draft]{agujournal2019}
\usepackage{url} 
\usepackage{lineno}
\usepackage[inline]{trackchanges} 
\usepackage{soul}
\usepackage{amsmath}
\usepackage{amssymb}
\justifying
\draftfalse

\journalname{Geophysical Research Letters}

\begin{document}

%
\title{Are Dipolarization Fronts a Typical Feature of Magnetotail Plasma Jets Fronts?}

%
\authors{L.~Richard\affil{1,2}, 
        Yu.~V.~Khotyaintsev\affil{1}, 
        D.~B.~Graham\affil{1}, 
        C.~T.~Russell\affil{3}}

\affiliation{1}{Swedish Institute of Space Physics, Uppsala, Sweden}
\affiliation{2}{Space and Plasma Physics, Department of Physics and Astronomy, Uppsala University, Sweden}
\affiliation{3}{Department of Earth Planetary and Space Sciences, University of California Los Angeles, Los Angeles, CA, USA}

\correspondingauthor{Louis Richard}{louis.richard@irfu.se}

%
\begin{keypoints}
\item The magnetic field structures at magnetotail jet fronts are statistically investigated using a database of 2394 BBFs.
\item Solitary sharp large-amplitude dipolarization fronts are found at only 10\% of the magnetotail jet fronts.
\item Complex ``turbulent" magnetic field structures are found at 32\% of the magnetotail jet fronts.
\end{keypoints}

%
\begin{abstract}
Plasma jets are ubiquitous in the Earth's magnetotail. Plasma jet fronts (JFs) are the seat of particle acceleration and energy conversion. JFs are commonly associated with dipolarization fronts (DFs) representing solitary sharp and strong increases in the northward component of the magnetic field. However, MHD and kinetic instabilities can develop at JFs and disturb the front structure which questions on the occurrence of DFs at the JFs. We investigate the structure of JFs using 5 years (2017-2021) of the Magnetospheric Multiscale observations in the CPS in the Earth's magnetotail. We compiled a database of 2394 CPS jets. We find that about half (42\%) of the JFs are associated with large amplitude changes in $B_z$. DFs constitute a quarter of these large-amplitude events, while the rest are associated with more complicated magnetic field structures. We conclude that the ``classical" picture of DFs at the JFs is not the most common situation.
\end{abstract}

\section*{Plain Language Summary}
We statistically investigate the magnetic field structure associated with the leading edge of fast plasma flows (jets), in the nightside region of the Earth's magnetosphere. Owing to its stretched magnetic field configuration, this region is called the magnetotail. The magnetotail plasma jets transport magnetic flux in the form of structures carrying an intense dipole-like magnetic field. It is commonly assumed that the leading edge of the structures convected by the plasma jet is a sharp change of the magnetic field configuration from a stretched to dipole-like configuration called a dipolarization front. However, the plasma jet provides a source of free energy that dissipates through various instabilities which grow at the leading edge of the plasma jet. Hence, we suspect that the sharp vertical boundary is often strongly perturbed. Using the Magnetospheric Multiscale spacecraft observations, we find that the dipolarization fronts are seen in only 10\% of the plasma jet fronts. On the other hand, we found that 32\% of the plasma jet fronts are associated with magnetic field structures, which are more complex than dipolarization fronts. Our result questions the commonly accepted picture of dipolarization fronts as the leading edge of magnetotail plasma jets.

\section{Introduction}
Magnetotail bursty bulk flows (BBFs) \cite{baumjohann_characteristics_1990} are transient (10-100 s) fast plasma jets in the central plasma sheet (CPS). Earthward BBFs are responsible for the majority of the plasma and magnetic flux transport from the magnetotail to the inner magnetosphere \cite{angelopoulos_statistical_1994}.  The jet fronts (JFs) are often accompanied by a strong and sharp ion-scale magnetic field structures called dipolarization fronts (DFs) \cite{nakamura_motion_2002,ohtani_temporal_2004,runov_themis_2011}. 
DFs are identified in spacecraft data as solitary sharp and strong increases in the northward component of the magnetic field $B_z$ \cite{fu_occurrence_2012,runov_themis_2011}. Upstream of the DF, the magnetic field configuration is that of the unperturbed CPS with a small $B_z$ normal to the magnetotail current sheet. DFs are often preceded by a depletion of $B_z$ before the sharp increase of $B_z$ at the front \cite{runov_themis_2009,schmid_statistical_2011}. Following the front $B_z$ slowly $20~\textrm{s}$ decreases to its initial upstream value \cite{runov_themis_2011}. DFs are ion-inertial-length-scale \cite{schmid_statistical_2011} boundaries separating low temperature dense plasma in the preexisting CPS from hotter tenuous plasma \cite{runov_themis_2011,khotyaintsev_plasma_2011}. In the cross-tail dawn-dusk direction DFs extend over $\sim 1 - 5~R_E$ \cite{sergeev_detection_1996,liu_cross-tail_2015}. DFs are thought to be a result of transient unsteady magnetic reconnection \cite{sitnov_dipolarization_2009} or detachment of interchange heads in the nonlinear stage of the kinetic ballooning interchange instability \cite{pritchett_kinetic_2010}. Because of their distinct signature, DFs and their contribution to the magnetotail energy budget have been extensively investigated using case studies \cite{fu_fermi_2011,artemyev_ion_2012,angelopoulos_electromagnetic_2013,khotyaintsev_energy_2017} and statistical studies \cite{runov_themis_2011,schmid_statistical_2011,fu_occurrence_2012,schmid_dipolarization_2019}. These studies focused on DFs, and thus omitted other JFs associated with more complicated magnetic field structures. Hence, a fundamental question arises: Are DFs the most common magnetic field structures associated with JFs? 

The aforementioned question is particularly relevant since various instabilities are found to grow at the JFs due to strong gradients in field and plasma parameters present there. Such instabilities can grow to large amplitudes and distort the JF structure \cite{balikhin_fine_2014}. In particular, perturbations of the JFs can result from the MHD interchange mode \cite{guzdar_simple_2010,lapenta_self-consistent_2011}, the kinetic ballooning interchange instability \cite{pritchett_kinetic_2010}, the lower hybrid drift instability \cite{divin_lower_2015,pan_rippled_2018} and the pressure-anisotropy-driven mirror mode \cite{zieger_jet_2011} and firehose \cite{hietala_ion_2015} instabilities. We use the Magnetospheric Multiscale (MMS) data \cite{burch_magnetospheric_2016} to characterize the magnetic structure of the JFs.

\section{Data Sets}
We search for BBFs from the five MMS magnetotail seasons: 4 May 2017 to 30 August 2017, 25 May 2019 to 28 September 2019, 18 June 2019 to 5 October 2019, 27 June 2020 to 9 October 2020 and 26 June 2021 to 13 October 2021. We use the Fast Survey data. Among all the instruments onboard the MMS spacecraft, we use the magnetic field measurements from the Flux Gate Magnetometer (FGM) \cite{russell_magnetospheric_2016} and the moments of the ion distribution measured by the Fast Plasma Investigation Dual Ion Spectrometer (FPI-DIS) \cite{pollock_fast_2016}. To account for the penetrating radiation, we correct the ion moments using the method described in \citeA{gershman_systematic_2019}. We note that the data is not uniformly distributed over the 5 years, due to the switch off of the FPI instruments during the Earth's shadow encounters, which were longer during the seasons 2019 to 2021. Finally, since the scale of the BBF channel is typically of the order of a few Earth radii \cite{nakamura_spatial_2004} which is much larger than the MMS spacecraft separation of $\Delta R \sim 10~\textrm{km}$, the spacecraft tetrahedron can be considered as one single spacecraft. Therefore, we use the magnetic field and the ion moments measured by MMS 1 only.

\section{Spatial Distribution and Duration of BBFs}

\subsection{Selection Criteria}
To detect BBFs in the magnetotail, we employ a selection criteria based on the Geocentric Solar Magnetospheric (GSM) Earthward component of the ion bulk velocity $V_{ix}$. To allow for comparison with previous studies, we use criteria similar to those used by \citeA{baumjohann_characteristics_1990} and \citeA{angelopoulos_statistical_1994}. The criteria are the following:

\begin{enumerate}
    \item Peak ion bulk velocity $|V_{ix}| \geq 300~\textrm{km}~\textrm{s}^{-1}$. 
    \item BBF time intervals are defined such that $|V_{ix}| \geq 100~\textrm{km}~\textrm{s}^{-1}$.
    \item Time records which satisfy condition 1 are clustered with those 60 s apart.
    \item Central Plasma Sheet (CPS) BBFs with $\beta_i > 0.5$ \cite{angelopoulos_statistical_1994}, where $\beta_i = P_i / P_{mag}$, $P_i$ is the ion plasma pressure, and $P_{mag} = B^2/2\mu_0$ is the magnetic pressure.
    \item To prevent selection of magnetosheath jets \cite{plaschke_jets_2018}, the region of interest is restricted to $|Y_{GSM}| \leq 12~R_E$.
\end{enumerate}

We note that our peak velocity threshold is similar to $|V_{\perp x}| \geq 300~\textrm{km}~\textrm{s}^{-1}$, with $V_{\perp x}$ the Earthward component of the velocity perpendicular to the local magnetic field, used by \citeA{ohtani_temporal_2004}. This threshold is lower than $|V_{ix}| \geq 400~\textrm{km}~\textrm{s}^{-1}$ used by \citeA{baumjohann_characteristics_1990} and \citeA{angelopoulos_statistical_1994}, so that the resulting occurrence rate of BBFs may differ from that found in these studies. 

\subsection{Results}
\begin{figure}[!ht]
    \centering
    \includegraphics[width=\linewidth]{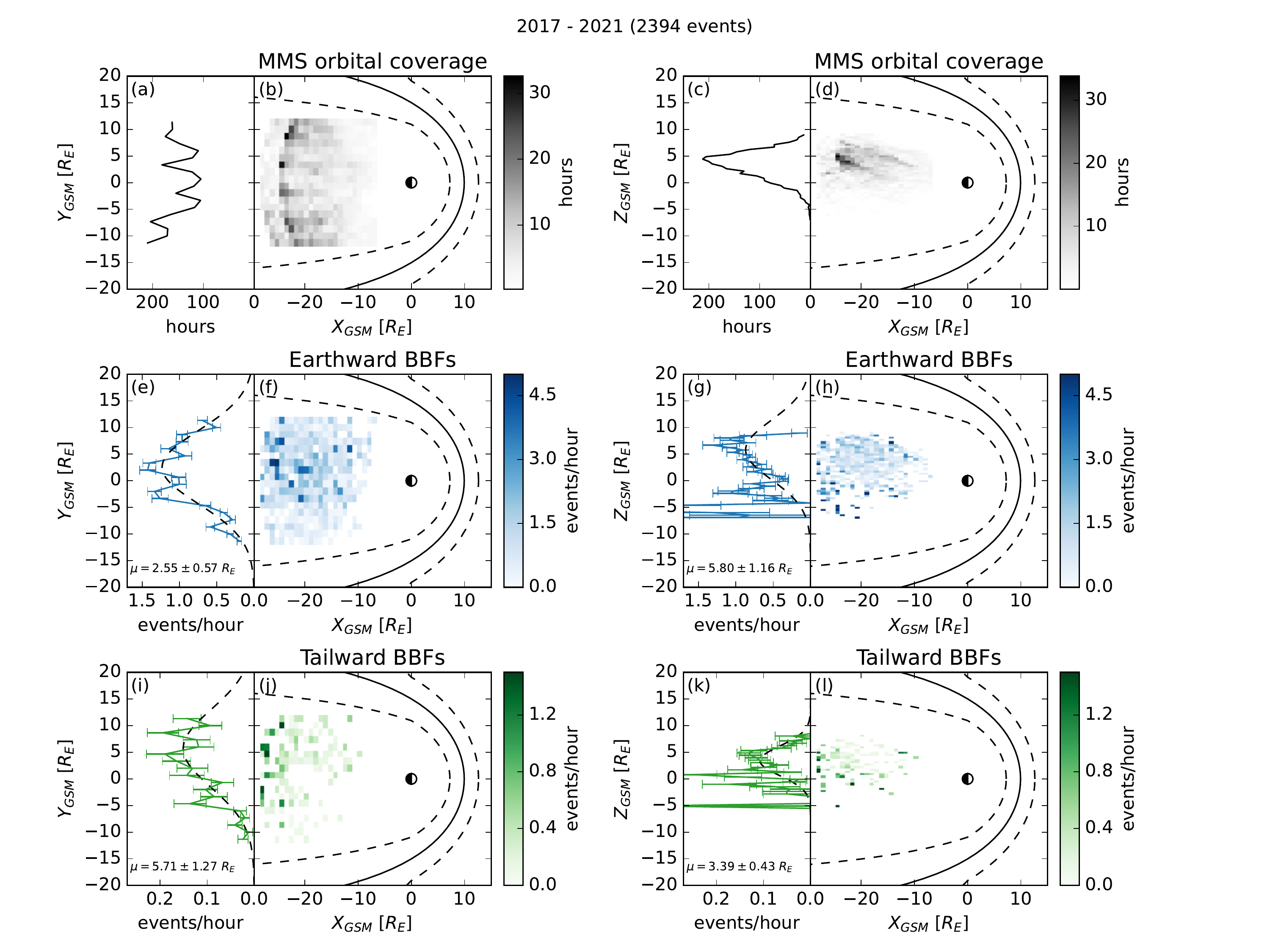}
    \caption{Spatial distributions of the MMS orbital coverage and Earthward and tailward BBFs. (a) Dawn-dusk, (b) equatorial plane, (c) north-south and (d) meridional plane projections of the distribution of the MMS orbital coverage. (e) Dawn-dusk, (f) equatorial plane, (g) north-south and (h) meridional plane projections of the distribution of the occurrence rate of Earthward BBFs. (i) Dawn-dusk, (j) equatorial plane, (k) north-south and (l) meridional plane projections of the distribution of the  occurrence rate of tailward BBFs. The black dashed lines in panels (e), (g), (i) and (k) are the Gaussian fits. The black lines in panels (a), (d), (f), (h), (j) and (l) are the average (solid line) and $1\sigma$ boundaries (dashed lines) of the statistical magnetopause from \citeA{shue_magnetopause_1998}. The error bars are obtained by propagating the standard Poisson statistical uncertainty of each bin count.}
    \label{fig:figure1}
\end{figure}

We find a total of 2394 BBFs in the CPS, with 2135 Earthward and 259 tailward BBFs. We note that 1796 of the BBFs also satisfy $\max(|V_{ix}|) \geq 400~\textrm{km}~\textrm{s}^{-1}$ and only 1175 BBFs satisfy $\max(|V_{ix}|) \geq 500~\textrm{km}~\textrm{s}^{-1}$. To account for orbital coverage bias, we plot in Figure~\ref{fig:figure1} the projection of the MMS CPS orbital coverage (1st row) and the occurrence rate (number of events per bin / orbital hours per bin) of Earthward (2nd row) and tailward (3rd row) BBFs. The statistically relevant size of the bin $h_i$ along the $i$ GSM axis is estimated using a Freedman-Diaconis estimator $h_i=2 IQR_i / n^{1/3}$, where $IQR_i$ is the interquartile range of the spacecraft position, and $n=2394$ is the total number of BBFs. It yields $h_x=0.9~R_E$, $h_y=1.3~R_E$ and $h_z=0.5~R_E$. Finally, we fit the dawn-dusk (panels e and i) and north-south (panels g and k) projection of the occurrence rate distribution to a Gaussian distribution (black dashed lines) using a Levenberg-Marquardt method of minimization of the $\chi^2$ cost function.

The MMS CPS orbital coverage is strongly skewed toward the northern GSM hemisphere (Figure~\ref{fig:figure1}b), and increases with distance from Earth due to slower orbital velocity near the apogee. The equatorial projection of the orbital coverage (Figure~\ref{fig:figure1}a) is not uniformly distributed in magnetic local time (MLT). This somewhat surprising observation is due to rotation of the Earth's magnetic dipole axis which lies in the $XZ$ GSM plane. 

The dawn-dusk (Figure~\ref{fig:figure1}e) and north-south (Figure~\ref{fig:figure1}g) distributions of the Earthward BBFs show clear Gaussian shapes centered at $\langle Y_{GSM}\rangle = 2.55\pm 0.57~R_E$ ($\sigma (Y_{GSM}) = 6.87\pm 0.59~R_E$) and $Z_{GSM} = 5.80\pm 1.16~R_E$ ($\sigma(Z_{GSM}) = 5.55\pm 1.28~R_E$) respectively. The dawn-dusk (Figure~\ref{fig:figure1}i) and north-south (Figure~\ref{fig:figure1}k) distributions of the tailward BBFs show, similar to that of Earthward BBFs, clear Gaussian shapes centered at $\langle Y_{GSM}\rangle = 5.71\pm 1.27~R_E$ ($\sigma (Y_{GSM}) = 7.34\pm 0.96~R_E$) and $Z_{GSM} = 3.39\pm 0.43~R_E$ ($\sigma(Z_{GSM}) = 2.79\pm 0.52~R_E$) respectively. Since during the MMS magnetotail phase (northern hemisphere summer) the neutral sheet (NS) is statistically located north of the GSM equatorial plane, the northward skewness of the north-south distributions of the Earthward and tailward BBFs is attributed to an observational bias.  

\begin{figure}[!ht]
    \centering
    \includegraphics[width=\linewidth]{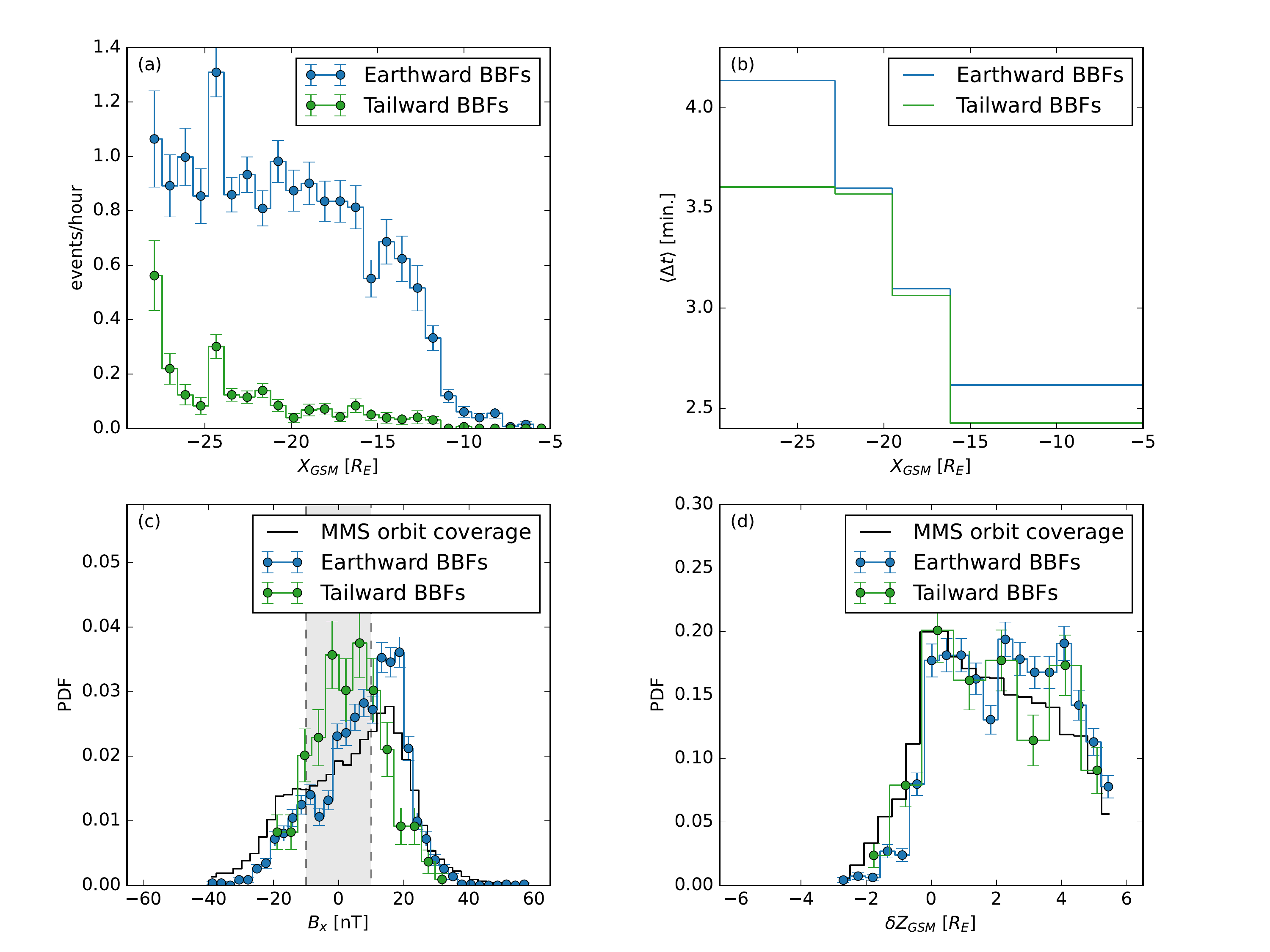}
    \caption{Distribution of Earthward (blue) and tailward (green) BBFs with respect to Earth and NS. (a) Earthward projection of the BBFs occurrence rate. (b) Average duration of the BBFs with respect to distance from Earth. (c) Probability Density Function of $B_x$ associated with BBFs. The black line indicates the PDF of $B_z$ measured by MMS during all magnetotail seasons. (d) Probability Density Function of $\delta Z_{GSM}$ associated with BBFs. The black line indicates the PDF of $\delta Z_{GSM}$ computed for MMS during all magnetotail seasons. The error bars are obtained by propagating the standard Poisson statistical uncertainty of each bin count.}
    \label{fig:figure2}
\end{figure}

Figure~\ref{fig:figure2}a shows the occurrence rate of Earthward (blue) and tailward (green) BBFs with respect to $X_{GSM}$. The occurrence of Earthward BBFs is constant for $X_{GSM} \lesssim -19~R_E$, consistent with the AMPTE/IRM and ISEE 2 statistical study by \citeA{angelopoulos_statistical_1994}, and decreases with distance from Earth in the $-19~R_E \lesssim X_{GSM} \lesssim -10~R_E$ region. At the same time, the duration of the Earthward and tailward BBFs decreases with distance from Earth (Figure~\ref{fig:figure2}b). The occurrence rate of the tailward BBFs is small but non-zero at $-5~R_E \geq X_{GSM} \gtrsim -15~R_E$, slightly increases at $X_{GSM}\sim -21~R_E$, is constant at $-21~R_E \gtrsim X_{GSM} \gtrsim -25~R_E$ and rises beyond $X_{GSM}\lesssim -25~R_E$.

To estimate the distance from the BBFs to the magnetotail NS, we plot the probability density function (PDF) of $B_x$, which is a proxy of the distance to the NS, as well as that of the distance $\delta Z_{GSM}$ to a model NS \cite{fairfield_statistical_1980} in Figure~\ref{fig:figure2}c and~\ref{fig:figure2}d respectively. We also plot the distribution of $B_x$ and $\delta Z_{GSM}$ measured during all magnetotail seasons (black). For the Earthward BBFs, the distributions of the two proxies overlaps with that of the background orbital coverage, suggesting that Earthward BBFs can be observed over a large north-south extent. For tailward BBFs, the distribution of $B_x$ is clearly more centered around the NS ($|B_x| = 0~\textrm{nT}$) and confined to $|B_x| < 10~\textrm{nT}$ which means that the tailward BBFs are likely to be observed over a narrow north-south extent.

\section{Occurrence of Dipolarization Fronts associated with BBFs}

\subsection{Selection Criteria}
To characterize the magnetic field structure at JFs, we employ the methods by \citeA{schmid_statistical_2011} [S11] and \citeA{fu_occurrence_2012} [F12]. The methods are applied to the magnetic field data in $180~s$ windows. Since the duration of a BBF is given by $|V_{ix}| > 100~\textrm{km}~\textrm{s}^{-1}$, the intervals shorter than 180 s are extended to 180 s, while longer intervals are split in 180 s intervals with a 90 s overlap. Since we are interested in the magnetic field structures at the JFs, i.e., upstream of the velocity peak, we use only the data before the peak. The following criteria from S11 are then applied to the 180 s windows:

\begin{enumerate}
    \item Significant change of $B_z$: $|\Delta B_z| = |\max (B_{z}) - \min (B_{z})| \geq 4~\textrm{nT}$.
    \item Significant change of inclination angle $\theta = \arctan \left ( B_z / \sqrt{B_x^2 + B_y^2} \right )$: $\Delta \theta = \theta(B_z=\max (B_z)) - \theta(B_z=\min (B_z)) \geq 10^\circ$.
    \item Inclination angle close to the dipolar configuration: $\max (\theta) \geq 45^\circ$.
\end{enumerate}
These criteria were originally designed to find DFs. However, the superposed epoch analysis in S11 shows that the median of the superposed $B_z$ does not show a distinct sharp increase of $B_z$ expected for DFs \cite<e.g.,>{runov_themis_2011}. From which we conclude that the S11 criteria corresponds to broader class of large-amplitude changes of $B_z$ than just the DFs (e.g., plasmoids, DFs, current sheets, etc). Thus to identify the DFs we in addition employ the F12 method, which relies on fitting the $B_z$ increase by a hyperbolic tangent function: 

\begin{equation}
    B_{fit} = \frac{a}{2}\tanh \left ( \frac{t - t_{DF}}{b/2}\right) + \left ( c + \frac{a}{2}\right ),
    \label{eq:fit}
\end{equation}
where $t_{DF}$ is the time of the candidate DF. First, we determine $t_{DF}$ as the time of maximum of the time derivative of $B_z$ smoothed using a 5th order Savitzky–Golay filter on a time scale of $0.5 f_{ci}^{-1}$. This is different from that done in F12 who used $t_{DF} = (t_{\mathrm{min}(B_z)} + t_{\mathrm{min}(B_z)}) / 2$. Then, we fit $B_z$ in the interval $[t_{DF} - 30~\textrm{s}, t_{DF} + 15~\textrm{s}]$.
Finally, we employ criteria on the amplitude $|a| \geq 4~\textrm{nT}$, the time scale $b \leq 8~\textrm{s}$ and the root-mean square residual $\sigma = \sqrt{\langle |B_z - B_{fit}|^2\rangle } \leq |a| / 2$. In contrast to the constant $\sigma \leq 2.5~\textrm{nT}$ used in F12, we use a criteria which depends on the amplitude of the $B_z$ jump.

\subsection{Results}
\begin{figure}[!ht]
    \centering
    \includegraphics[width=\linewidth]{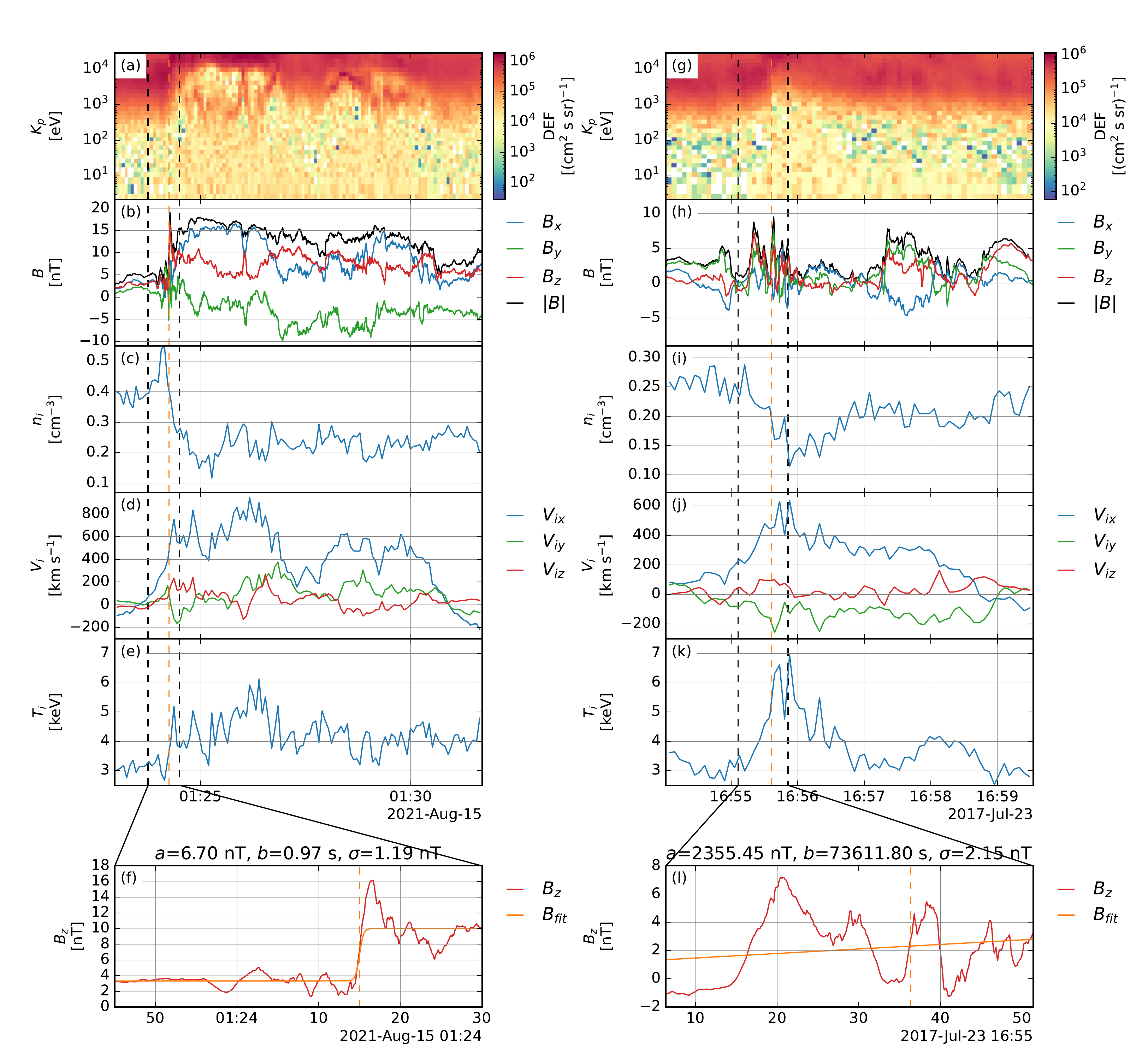}
    \caption{Examples of a ``classical" DF (left) and turbulent JF (right). (a), (g) Ion differential energy flux spectrum. (b), (h) Magnetic field in GSM coordinates. (c), (i) Ion number density. (d), (j) Ion bulk velocity in GSM coordinates. (e), (k) Ion temperature. (f), (l) $B_z$ used for fitting (red) and fitted magnetic field (orange). The orange dashed lines indicate the estimated times of the candidates dipolarization fronts (see text).}
    \label{fig:figure3}
\end{figure}

The left column of Figure~\ref{fig:figure3} shows an example of a BBF front with a DF, i.e., that satisfies both S11 and F12-based criteria. We observe that the time of the DF $t_{DF}$ (orange dashed line) matches the ion energy enhancement (panel a), the ion density density $n_i$ decrease (panel c) and the ion temperature $T_i$ increase (panel e), which indicates that our method provides an accurate estimate of $t_{DF}$. This case presents a solitary large amplitude sharp increase of $B_z$ and thus corresponds to the ``classical" picture of a DF \cite{runov_themis_2011}. The fitting procedure applied to $B_z$ (panel f) yields the amplitude $a=6.7~\textrm{nT}$, the time scale $b=0.97~\textrm{s}$ and $\sigma = 1.19~\textrm{nT}$ that satisfy the F12-based criteria, so this JF is classified as a DF.

The right column of Figure~\ref{fig:figure3} presents an example of a BBF front which satisfies the S11 criteria but not the F12-based criteria. This JF is associated with multiple ion-scale magnetic field structures (panel h). This event was first reported by \citeA{alqeeq_investigation_2022} to show the non-homogeneity of the energy conversion due to electron-scale substructures attributed to lower-hybrid drift waves.  Similar to the other example, $t_{DF}$ is accurately estimated as it corresponds to the ion energy enhancement (panel g), the ion density $n_i$ decrease (panel i) and the ion temperature $T_i$ increase (panel k). However, the F12 fitting procedure applied to the $B_z$ component of the complex magnetic field structures (panel l) yields the amplitude $a=2355~\textrm{nT}$, the time scale $b=73611~\textrm{s}$ and $\sigma = 2.15~\textrm{nT}$ which clearly do not satisfy the F12-based criteria. So this JF example, which satisfies the S11 criteria but not the F12-based criteria, is not classified as a DF. 

We obtain a total of 1013 BBFs which satisfy the S11 criteria, with 904 Earthward BBFs and 109 tailward BBFs. Among the 1013 BBFs that satisfy S11 criteria, we find that only 247 also satisfy F12-based criteria with 238 Earthward BBFs and 9 tailward BBFs. Since JFs that satisfy the S11 but not F12-based criteria are associated with magnetic field structures more complex than DFs, we refer to them as ``turbulent" JFs. There are in total 766 of such BBFs. The remaining 1381 of the BBFs do not satisfy S11 criteria. These BBFs have relatively small changes of $B_z$ so we refer to them as quiet JFs. We summarize the occurrence of each type of JFs (quiet, DF, and turbulent JFs) in Table \ref{tab:occurrences}. The percentages are given in terms of total number of Earthward, tailward and all BBFs.
About half (42\%) of the JFs are associated with significant increase in $B_z$, $|\Delta B_z| > 4~\textrm{nT}$. Among these events, turbulent JFs are 3 times more common than JFs with a DF.

\begin{table}
    \caption{Occurrence of quiet, DF associated and turbulent BBFs}
    \centering
    \begin{tabular}{l r r r}
        \hline
        Criteria & Earthward BBFs & Tailward BBFs & All BBFs\\
        \hline
        Quiet JFs $\overline{\textrm{S11}}$ & 1231 (58\%) & 150 (58\%) & 1381 (58\%) \\
        DFs $\textrm{S11}\cap \textrm{F12}$ & 238 (11\%) & 9 (3\%) & 247 (10\%) \\
        Turbulent JFs $\textrm{S11} - \textrm{F12}$ & 666 (31\%) & 100 (39\%) & 766 (32\%) \\
        Total & 2135 (100\%) & 259 (100\%) & 2394 (100\%) \\
        \hline
    \end{tabular}
    \label{tab:occurrences}
\end{table}

\section{Discussion}
Using the MMS observations, we found 2394 BBFs in the CPS. We find that the distribution of both Earthward and tailward BBFs is skewed toward the the dusk side. Similar dawn-dusk asymmetry was observed by \citeA{mcpherron_characteristics_2011} using THEMIS, who showed that Earthwards BBFs are predominantly found in the pre-midnight sector. \citeA{lu_hall_2016} suggested that the dawn-dusk asymmetry of the magnetotail results from a stronger Hall electric field on the dusk side which is due to higher perpendicular ion temperature, thinner current sheet and small $B_z$. We find that the occurrence rate of the Earthward BBFs decreases with distance to Earth, suggesting that Earthward BBFs are decelerated in their course to the Earth leading to a threshold effect in the selection of BBFs. The deceleration of the BBFs during their Earthward convection has been attributed to deflection due to diamagnetic drift which results from magnetic pressure increase \cite{angelopoulos_characteristics_1993}, the Joule dissipation \cite{zhang_bbf_2020}, or energy loss through emission of kinetic Alfvén waves \cite{angelopoulos_plasma_2002,chaston_correction_2012}. We showed that the duration of the Earthward BBFs decreases with distance to Earth, which can be due to our choice of $|V_{ix}| > 100~\textrm{km}~\textrm{s}^{-1}$ wide intervals, so that slower BBFs are artificially shortened. 

At distances where the Earthward BBFs decelerate, we showed a small but non-zero occurrence of tailward BBFs. Tailward BBFs at distances $-5~R_E \geq X_{GSM} \gtrsim -15~R_E$ can result from velocity shear/magnetic field line twist induced eddies at the edges of the Earthward flow channel \cite{birn_propagation_2004,keiling_substorm_2009,zhang_measurements_2019}, kinetic ballooning interchange instability \cite{pritchett_kinetic_2010}, or flow rebounds \cite{chen_theory_1999,panov_multiple_2010,ohtani_tailward_2009}. The latter mechanism was suggested by \citeA{ohtani_tailward_2009}, to be the most likely cause of tailward BBFs, which are often preceded with Earthward BBFs. However, here, we found that only 18 (1.7\%) tailward BBFs in the $-5~R_E \geq X_{GSM} \gtrsim -20~R_E$ are preceded (within 10 mins) by an Earthward BBF. This result suggests that the incident Earthward convected flux tube is non-specularly reflected from its equilibrium \cite{ohtani_tailward_2009}, or that the tailward BBFs are much slower than the incident Earthward BBFs resulting from heavy damping of the oscillations of the flux tube \cite{chen_theory_1999,panov_multiple_2010}.

Using the $B_z$ fitting based on the method by \citeA{fu_occurrence_2012}, we find that only 10\% of the the BBFs have DFs at their front. This result contradicts the superposed Epoch analysis carried out by \citeA{ohtani_temporal_2004} and \citeA{wiltberger_highresolution_2015} using Geotail observations and the Lyon-Fedder-Mobarry global MHD magnetosphere model. \citeA{ohtani_temporal_2004} and \citeA{wiltberger_highresolution_2015} showed that the magnetic field convected Earthward with the fast flows has a dipolar configuration at the JF. However, in their analysis \citeA{ohtani_temporal_2004} and \citeA{wiltberger_highresolution_2015} selected the BBFs based on ion velocity $V_{\perp x}$ perpendicular to the local magnetic field, which implicitly constrains the magnetic field to a dipolar configuration (large $B_z$), and therefore restrict the events to JFs with a DF. Our results suggest that the ``classical" picture of DF as a solitary sharp large amplitude increase of $B_z$ \cite{sitnov_dipolarization_2009,runov_themis_2011,fu_occurrence_2012} is not the most likely magnetic field structure associated with the JF.

Furthermore, using the selection criteria proposed by \citeA{schmid_statistical_2011} to detect magnetic field changes, we show that these criteria allow us to detect not only the sharp DF but also more complex magnetic field structures at the JF. In particular, we found that 32\% of the BBFs are associated with complex magnetic field structures other than DFs. The abundance of such complex structures at the JF over DF can be explained by the wide variety of instabilities which can develop at fronts of  BBFs, such as the MHD interchange instability \cite{guzdar_simple_2010,lapenta_self-consistent_2011}, the kinetic ballooning interchange instability \cite{pritchett_kinetic_2010}, the lower hybrid drift instability \cite{divin_lower_2015} and the pressure anisotropy driven mirror mode \cite{zieger_jet_2011} and firehose \cite{hietala_ion_2015} instabilities. The growth of the aforementioned instabilities results in electromagnetic fluctuations modifying the JF structure. These fluctuations in turn dissipate into particle energy through plasma heating \cite{angelopoulos_electromagnetic_2013,khotyaintsev_energy_2017} and particle acceleration \cite{khotyaintsev_plasma_2011,greco_role_2017}. The energy cascade of the free energy injected at the jet scale to the kinetic scales, suggests that turbulence develops within the BBFs \cite{voros_bursty_2006,huang_observations_2012,pucci_properties_2017,bandyopadhyay_observation_2020}. Our results suggest that BBFs accompanied with ``turbulent" JF with magnetic field oscillations are 3 times more likely than BBFs associated with DFs.

\section{Conclusion}
We compiled a database of 2394 plasma sheet jets (BBFs), which show statistical properties similar to the ones known from earlier studies \cite{baumjohann_characteristics_1990,angelopoulos_statistical_1994,mcpherron_characteristics_2011}. We characterize the structure of the jet fronts (JFs) using the criteria by \citeA{schmid_statistical_2011} and the method introduced by \citeA{fu_occurrence_2012} with a new technique based on the magnetic field gradient to estimate the time of the candidate DF. We find that only 10\% of the JFs are associated solitary sharp and strong dipolarization of the magnetic field commonly refered to as a dipolarization front. About half (42\%) of jet fronts (JFs) are associated with large variations of the north-south component of the magnetic field $B_z$, $|\Delta B_z| > 4~\textrm{nT}$. DFs constitute a quarter of these events, while the rest are associated with more complicated magnetic field structures i.e., ``turbulent''. Our results indicate that the classical picture of a DF as the magnetic field structure associated with the JF is not the most commonly occurring magnetic field structure, which points at instabilities to drive ion to electron scale perturbations modifying the structure of the magnetotail plasma jet fronts \cite{balikhin_fine_2014}.

%
\section{Open Research}
All data used in this paper are publicly available from the MMS Science Data Center \url{https://lasp.colorado.edu/mms/sdc/}. Data analysis was performed using the pyrfu analysis package available at \url{https://pypi.org/project/pyrfu/}. The codes to reproduce the figures in this paper are available at \url{https://github.com/louis-richard/bbfstats}. The compiled BBFs database and additional data are available at \url{https://doi.org/10.5281/zenodo.7009706}.

\acknowledgments
We thank the entire MMS team and instrument PIs for data access and support. LR thanks A. Lalti, C. Norgren and the International Space Science Institute (ISSI) working group on ``Magnetotail Dipolarizations: Archimedes Force or Ideal Collapse?" for valuable discussions. This work is supported by the Swedish National Space Agency grant 139/18.

%
\bibliography{main}
\newpage

%

\end{document}